\documentstyle[aps]{revtex}
\begin{document}
\def\be{\begin{equation}}
\def\ee{\end{equation}}
\def\bea{\begin{eqnarray}}
\def\eea{\end{eqnarray}}
\renewcommand{\thefootnote}{\fnsymbol{footnote}}
\def\nin{\noindent}

\twocolumn[\hsize\textwidth\columnwidth\hsize\csname@twocolumnfalse%
\endcsname

\title{Non perturbative renormalization group potentials and quintessence }
\author{Vincenzo Branchina}
\address{Laboratoire de Physique Th\'eorique, Universit\'e Louis Pasteur,}
\address{3-5, rue de l'Universit\'e, F-67084, Strasbourg Cedex, France}

\date{\today}
\maketitle
\draft
\begin{abstract}
New solutions to the non preturbative renormalization group equation for the 
effective action of a scalar field theory in the Local Potential Approximation  
having the exponential form $e^{\pm\phi}$ are found. This result could be relevant 
for those quintessence phenomenological models where this kind of potentials are 
already used, giving them a solid field theoretical derivation. Other non perturbative 
solutions, that could also be considered for the quintessence scenario, are also found. 
Apart from this particular cosmological application, these results could be relevant for 
other models where scalar fields are involved, in particular for the scalar sector of 
the standard model.  
\end{abstract} 

\pacs{pacs 98.80.Cq, 11.10.Hi , 11.10.Gh \hfill}
]
One of the most challenging problems in fundamental physics has been 
the search for a theoretical argument, hopefully a symmetry, that could explain the 
vanishing of the cosmological constant\cite{wein}. Recent observations from high 
redshift supernovae\cite{perl}, combined with the data on the fluctuation of 
the cosmic microwave background\cite{blanch}, have changed our perspective. 
Apparently $\Omega_M$, the ratio of the 
barionic and cold dark matter density to the critical density, is about $\frac{1}{3}$. 
This means that either the universe is open or the missing energy is provided by some 
new form of matter.  
The simplest candidate is a cosmological constant term. Alternative models 
where the missing energy is given by a scalar field slowly rolling down its 
potential, the ``quintessence"\cite{stein}, have recently attracted lot of attention. 

For a truly constant vacuum energy term the ``old" problem of explaining the
vanishing of the cosmological constant is replaced by the equally difficult one of
explaining why it has the small observed value of about $(3\times 10^{-3} ev)^4$.
In the quintessence scenario the equivalent problem  is the
so called ``coincidence problem". The matter and the scalar fields
evolve differently but we observe today an order of magnitude ``coincidence" between the matter
energy density $\Omega_M$ and the quintessence energy density $\Omega_\phi$ that requires a
fantastic fine tuning of the initial conditions. The notion of ``tracker field" \cite{stein2},
a quintessence scalar field that evolves to an attractor solution during its rolling down,
has been introduced to circumvent this problem. For a wide range of initial conditions the
attractor is stable. An explanation for the coincidence can be obtained this way and  
this has been advocated as an argument in favour of the quintessence scenario\cite{stein2}. 
Recently another interesting argument has been given in the framework of the brane-world 
picture\cite{stein3}. 
 
Even restricting ourselves to tracker fields there is still a large amount of 
arbitrariness in the possible form of this potential. Different proposals
\cite{stein,stein2,fri,joy,joy2,sko,barr,sah} have been essentially based on their capability 
to reproduce the observational data. Also interesting attempts have been done
\cite{bin,mas,bar,brax,ben} to derive its form from particle physics models, but they still leave 
the door open to several possibilities and it is not possible to discriminate between the
different proposals.

In this Letter I show that the renormalization group equation for the effective action 
of a single component scalar field theory in the Local Potential Approximation (LPA)
possesses non perturbative solutions in addition to the well known perturbative one.
These are of the form of exponential potentials that are among the favoured candidates in the 
quintessence scenario\cite{stein,stein2,stein3,joy2,barr}. I think that this result can give a
strong motivation to these potentials. I will comment more on this point later.

The exact renormalization group equation for the effective action has been found 
in Ref.\cite{weg}. By considering only the potential term, an approximation to this equation 
(the LPA) is obtained\cite{nicol}, and for a single component scalar field theory in $d=4$ 
dimensions it reads: 

\be\label{lpa}
k\frac{\partial}{\partial k}U_k(\phi)=-
\frac{k^4}{16\pi^2} ln \Bigl(\frac{k^2 + U_k^{''}(\phi)}{k^2 + m^2}\Bigr).
\ee

\noindent
Here $k$ is the current scale, $m^2$ is a constant with dimension $({\rm mass})^2$, $U_k$ is
the potential at the scale $k$ and the $'$ means derivative with respect to 
the field $\phi$.

\noindent
Eq.(\ref{lpa}) is a non perturbative evolution equation for $U_k$.
It is immediate to see that the k-independent solution to Eq.(\ref{lpa}), i.e. the fixed 
point potential, is the trivial gaussian one:

\be\label{gp}
U_{f}(\phi)=\frac{1}{2} m^2\phi^2 + \alpha\phi +\beta.
\ee

One simple way to show that  Eq.(\ref{lpa}) admits the well known perturbative solution 
is as follows. Let's consider first a small deviation of the potential 
$U_k(\phi)$ around $U_f(\phi)$,

\be\label{dev}
U_k(\phi)=U_f(\phi) + \delta U_k(\phi).
\ee

\noindent
We develop now the logarithm in Eq.(\ref{lpa}) in powers of $\delta U_k$
and expand $\delta U_k$ in powers of the field (for the sake
of simplicity we consider a potential with the $Z(2)$ symmetry $\phi \to -\phi$)

\be\label{del}
\delta U_k(\phi)= \frac{\lambda_2(k)}{2}\phi^2 +\frac{\lambda_4(k)}{4!}\phi^4+\cdots.
\ee

\noindent
At any order in $\delta U_k$ we obtain this way  an infinite system of equations 
for the coupling constants. By truncating this system to the first two equations 
for $\lambda_2(k)$ and $\lambda_4(k)$ and solving iteratively up to the second 
order in $\delta U_k$ \cite{com1} we get the known perturbative one loop RG flow
for the coupling constants. Extrapolating down to $k=0$, this flow identifies the gaussian 
fixed point as infrared stable, i.e. we have the standard result about the triviality of 
the theory. 
 
We ask now the question about the possibility of having non perturbative solutions. 
The only k-independent solution to Eq.(\ref{lpa}), i.e. the
only fixed point potential, is the gaussian one found in Eq.(\ref{gp}). 

We want to linearize now Eq.(\ref{lpa}) around $U_f$ and look for a {\it small}
but {\it non perturbative} $\delta U_k$. We have (for notational
simplicity I write $U_k$ rather than $\delta U_k$):

\be\label{lin}
k\frac{\partial}{\partial k}U_k(\phi)=-
\frac{k^4}{16\pi^2}\frac{1}{k^2 + m^2} U_k^{''}(\phi).
\ee

\noindent
There is a class of non perturbative solutions to Eq.(\ref{lin}) that is very easy to 
find. Let's seek for solutions of the form:

\be\label{fac}
U_k(\phi)=f(k)g(\phi).
\ee

\noindent
Once inserted in Eq.(\ref{lin}), the ansatz (\ref{fac}) gives:

\be\label{sy1}
\frac{d^2 g(\phi)}{d \phi^2}-\frac{1}{\mu^2}g(\phi)=0 ~~~~~~~~~~~~~~~~~
\ee
\be\label{sy2}
\frac{d f(k)}{d k}+\frac{1}{16\pi^2\mu^2}\frac{k^3}{k^2 + m^2}f(k)=0,
\ee

\noindent 
where $\mu^2$ is any constant with dimension $(\rm mass)^2$ that allows to separate 
Eq.(\ref{lin}) in the two Eqs. (\ref{sy1}) and (\ref{sy2}).  

Solving now these equations is a simple exercise. 
For positive values of the constant $\mu^2$ the solutions to Eq.(\ref{lin}) have the form 

\be\label{so1}
U_k(\phi)=M_1^4~e^{-\frac{1}{32\pi^2\mu_1^2}
\bigl(k^2-~m^2{\rm ln}\frac{k^2 + m^2}{m^2}\bigr)}e^{-\frac{\phi}{\mu_1}} ~~
\ee
\be\label{so2}
U_k(\phi)=M_2^4~e^{-\frac{1}{32\pi^2\mu_2^2}
\bigl(k^2-~m^2ln\frac{k^2 + m^2}{m^2}\bigr)}e^{+\frac{\phi}{\mu_2}},
\ee

\noindent 
for arbitrary values of the mass dimension constants $M_i$ and $\mu_i$.

For negative values of $\mu^2$ (calling  $\bar\mu^2=-\mu^2 >0$), 
the solutions to Eq.(\ref{lin}) have the form:

\be\label{so3}
U_k(\phi)=M_3^4~e^{+\frac{1}{32\pi^2\bar\mu_3^2}
\bigl(k^2-~m^2ln\frac{k^2 + m^2}{m^2}\bigr)}{\rm cos}\Bigl(\frac{\phi}{\bar\mu_3}\Bigr)
\ee
\be\label{so4}
U_k(\phi)=M_4^4~e^{+\frac{1}{32\pi^2\bar\mu_4^2}
\bigl(k^2-~m^2ln\frac{k^2 + m^2}{m^2}\bigr)}{\rm sin}\Bigl(\frac{\phi}{\bar\mu_4}\Bigr).
\ee

For solutions of the kind (\ref{so1}) and (\ref{so2}) the gaussian potential $U_f(\phi)$ 
is an ultraviolet fixed point. From that we immediately understand the completely different
nature of these solutions with respect to the perturbative one. 

As already mentioned these exponential potentials $e^{\pm\phi}$,
as well as linear combinations of them, are among the favourite candidates for the quintessence 
scenario\cite{stein,stein2,barr}.

The attempts that have been done to derive different forms of quintessence potentials 
all started from some sort of ``fundamental" higher 
energy model. For example, inverse power-law potentials have been motivated from Supersymmetric 
QCD \cite{bin,mas}. For another derivation of the same kind of potentials see\cite{bar}. 
Exponential potentials of the form 
found above arise naturally in several higher energy/higher dimensional theories\cite{co}.
As we simply don't know the theory that describes our world at very high energies,
the use of phenomenologically motivated potentials is certainly well justified. In that respect 
indications coming directly from the effective theory of the quintessence field should be
considered as very welcome.
Actually, whatever the structure of the higher energy/higher dimensional theory, i.e.
whatever the fundamental origin of the scalar quintessence field is, the ``low 
energy" effective theory for this field should be very well described by  
Eq.(\ref{lpa}). This happens
because the higher energy degrees of freedom decouple from the quintessence field.
There could still be the problem of the coupling of this field to
ordinary matter, i.e. to ordinary standard model fields. As the long range forces that these
couplings would generate are not observed, we can suppose that they are suppressed 
through some mechanism as for instance the one proposed in Ref.\cite{car}. In this case     
the renormalization group Eq.(\ref{lpa}) gives the flow equation for the quintessence field
irrespectively of the nature of the higher energy theory and the above results 
(\ref{so1}) and (\ref{so2}) give a solid motivation from the ``low energy side" 
to the phenomenological exponential potentials $e^{\pm\phi}$.

Concerning the solutions of the kind (\ref{so3}) and (\ref{so4}), we see that 
the gaussian potential $U_f(\phi)$ is neither an infrared nor an ultraviolet fixed 
point for them, even though we can multiply these solutions times a small dimensionless 
constant so that they still make sense as linearized solutions of the Eq.(\ref{lpa}).
I want to mention here that cosine potentials have also been considered as possible 
quintessence candidates\cite{fri,stein}. 

We want to seek now for other non perturbative solutions. 
Actually there is at least  another class of such solutions that can be easily found. 
To see that, we switch first to the dimensionless form of 
Eq.(\ref{lpa}). If we define the dimensionless field $\varphi$, the dimensionless scale 
parameter $t$ and  the dimensionless potential $v(\varphi,t)$ from

\be\label{dles}
\varphi=\frac{1}{4\pi}\frac{\phi}{k},~~~~
t={\rm ln}\frac{\Lambda}{k}, ~~~~
U_k(\phi)=\frac{k^4}{16\pi^2} v(\varphi,t),
\ee 

\noindent
where $\Lambda$ is a boundary value for $k$, Eq.(\ref{lpa}) becomes:

\be\label{dl}
\frac{\partial v}{\partial t} + 
\varphi\frac{\partial v}{\partial \varphi} - 4 v  =
ln \Biggl(\frac{1 +\frac{\partial^2 v}{\partial \varphi^2}}{1 +\frac{m^2}{\Lambda^2}e^{2t}}\Biggr).
\ee

The dimensionless potential $v_{f}(\varphi,t)$ that corresponds to the gaussian potential 
$U_f(\phi)$ of Eq.(\ref{gp}) is:

\be\label{gpdl}
v_{f}(\varphi,t)=\frac{1}{2}\frac{m^2}{\Lambda^2}e^{2t}\varphi^2 + 
\frac{4\pi\alpha}{\Lambda^3}e^{3t}\varphi 
+\frac{16\pi^2\beta}{\Lambda^4}e^{4t},
\ee

\noindent
and solves the equation:

\be\label{om}
\frac{\partial v}{\partial t} + 
\varphi\frac{\partial v}{\partial \varphi} - 4 v  = 0.
\ee

\noindent
Actually Eq.(\ref{dl}) is more often written as\cite{hasen}:

\be\label{dle}
\frac{\partial v}{\partial t} + 
\varphi\frac{\partial v}{\partial \varphi} - 4 v  =
ln \Biggl(1 +\frac{\partial^2 v}{\partial \varphi^2}\Biggr),
\ee

\noindent
i.e. by setting $m^2=0$. This simply corresponds to choose a massless fixed point potential 
and from now on we also restrict ourselves to this case.  
We consider now a small fluctuation around the fixed point potential,

\be\label{devi}
v(\varphi,t)=v_f(\varphi,t) + \delta v(\varphi,t),
\ee

\noindent
and linearize Eq.(\ref{dle}) around $v_f$ (again we write $v(\varphi,t)$ rather than 
$\delta v(\varphi,t)$) to get:        

\be\label{dlin}
\frac{\partial v}{\partial t} + 
\varphi\frac{\partial v}{\partial \varphi} - 4 v  =
\frac{\partial^2 v}{\partial \varphi^2}.
\ee

\noindent
By following the same strategy as before we look for solutions to Eq.(\ref{dlin}) with 
factorized $t$ and $\varphi$ dependence:

\be\label{ans}
v(\varphi,t)=f(t)g(\varphi).
\ee

\noindent
Inserting the ansatz (\ref{ans}) in Eq.(\ref{dlin}) we have:
 
\be\label{sys1}
\frac{d^2 g(\varphi)}{d \varphi^2} -\varphi\frac{d g(\varphi)}{d \varphi} 
+ 4 g(\varphi)=\alpha g(\varphi)
\ee
\be\label{sys2}
\frac{d f(t)}{d t}=\alpha f(t), ~~~~~~~~~~~~~~~~~~~~~~~~~~~
\ee

\noindent 
where $\alpha$ is an arbitrary dimensionless constant that allows to separate 
Eq.(\ref{dlin}) in the two Eqs. (\ref{sys1}) and (\ref{sys2}).

\noindent
Equation (\ref{sys2}) is trivially solved and gives:

\be
f(t)= A e^{\alpha t},
\ee

\noindent
where $A$ is the integration constant.

\noindent
The solution to Eq.(\ref{sys1}) can be found by series. Writing

\be \label{ser}
g(\varphi)=\sum_{n=0}^{\infty} c_n \varphi^n,
\ee

\noindent
inserting (\ref{ser}) in Eq.(\ref{sys1})
and exploiting the recurrence relations between the coefficients $c_n$, we get
the two linearly independent solutions ($a=\alpha-4$):

\be\label{se}
g_1(\varphi)=c_0\Biggl(1+ \sum_{n=1}^{\infty}
\frac{\Pi_{j=1}^n(a+2j-2)}{(2n)!}\varphi^{2n}\Biggr) ~~~~
\ee
\be\label{sep}
g_2(\varphi)=c_1\Biggl(\varphi+\sum_{n=1}^{\infty}\frac{\Pi_{j=1}^n(a+2j-1)}
{(2n+1)!}\varphi^{2n+1}\Biggr)
\ee

\noindent
where I have explicitly kept $c_0$ and $c_1$, the first two coefficients in Eq.(\ref{ser}), as  
integration constants.

After some trivial algebra we see that Eqs.(\ref{se}) and (\ref{sep}) can be written in
terms of confluent hypergeometric functions and the general solution to Eq.(\ref{sys1})
takes the compact form

\be\label{so}
g(\varphi)=c_0 {\it M}\Biggl(\frac{a}{2},\frac{1}{2},\frac{\varphi^2}{2}\Biggr) 
+c_1\varphi {\it M}\Biggl(\frac{a+1}{2},\frac{3}{2},\frac{\varphi^2}{2}\Biggr).
\ee

\noindent
We recall here that the confluent hypergeometric function $M(a,b,x)$ is defined as 

\be
M(a,b,x)=1+\frac{a}{b}x+\frac{1}{2!}\frac{a(a+1)}{b(b+1)}x^2+\cdots.
\ee

\noindent
Of course once dimensionless solutions to Eq.(\ref{dlin}) are known, it is a trivial exercise
to reconstruct dimensionfull potentials from Eq.(\ref{dles}). Some of the potentials 
obtained this way could also be considered as quintessence candidates.

It is worth to compare the general solution (\ref{so}) to Eq.(\ref{sys1})
with a non perturbative result that has been obtained some years ago by Halpern and 
Huang\cite{hua1,hua2} following a different but essentially equivalent approach. 
Searching for alternatives 
to the trivial $\phi^4$ theory, the authors expanded the potential in even powers of 
the field and derived an infinite system of differential equations for the coupling constants. 
Looking for new eigendirections in the parameter space, they actually ended with the solution 
$g_1(\varphi)$ to Eq.(\ref{sys1}). More precisely, 
after making the trivial changes to match the two different notations and restricting ourselves to 
consider the $N=1$ and d=4 case as in the present paper, we can immediately check that eq.(49) of 
Ref.\cite{hua2} coincides with the $g_1(\varphi)$ solution above. As these authors considered 
potentials containing only even powers of the field, obviously they could 
only get the solution $g_1(\varphi)$. 

At a first sight it could seem that the solution $g_2(\varphi)$ should be discarded
as it contains odd powers of the field and as such it is unbounded from below. 
We would  conclude in this case that the only physically acceptable general solution to 
Eq.(\ref{dlin}) is $g_1(\varphi)$, i.e. the Halpern-Huang result. But this is not always true.
It is immediate to see from Eq.(\ref{so}) that for all the positive or negative integer odd values 
of $\alpha$ such that $\alpha<4$, $g_2(\varphi)$ is a polynomial in $\varphi$ and 
it can be combined with $g_1(\varphi)$ to give a bounded from below potential. 
It is also easy to give examples where even when the hypergeometric function in $g_2(\varphi)$ keeps
all the infinite terms, still a linear combination of $g_1(\varphi)$ and $g_2(\varphi)$ 
gives a bounded from below 
potential. Take for instance the case $\alpha=5$ for which both $g_1$ and $g_2$ are not polynomials. 
From the asymptotic behaviour of the hypergeometric function we easily see that $c_0$ and 
$c_1$ can be chosen in such a way that the resulting potential is bounded from below. 
The class of physically acceptable potentials, that are solutions of the Eq.(\ref{sys1}),
is larger than that spanned by $g_1$ only.
   
I should mention at this point that an attempt to solve Eq.(\ref{dlin}) has been recently 
made in Ref.\cite{bo} where the solution $v(\varphi,t)=e^{5t}e^{\frac{\varphi^2}{2}}$ is
presented\cite{com3}.
The author says that this solution is asymptotically similar to those of Ref.\cite{hua1} 
but that the connection between the two results is unclear to him. 
From Refs.\cite{hua1,hua2} and from Eq.(\ref{so}) above (setting $c_1=0$) we immediately see that 
this solution is just the Halpern-Huang result for $\alpha=5$. It is also a trivial exercise to verify 
that the other solutions presented in\cite{bo}  (see Eq.(21) of that paper) are particular cases 
of the general solution, Eq.(\ref{so}), obtained for the integer values  $\alpha=6,7,8,\cdots$.

To summarize in this letter I have presented new solutions to the LPA of the exact renormalization
group equation for the effective action of a single component scalar field theory. These potentials
have the exponential form $e^{\pm\phi}$ and 
have been recently used in phenomenological quintessence models. 
As the effective theory for 
the quintessence field should be governed by Eq.(\ref{lpa}) whatever its higher energy origin, 
I argue that the fact that they arise as solutions to the renormalization group equation gives a 
{\it derivation} of these potentials that is alternative and complementary to those based on higher 
energy theories and should allow to discriminate between different proposals. 
I have also presented other solutions to the LPA renormalization group equation some of which 
already partially known\cite{hua1,hua2}.       

Apart from the application to the particular cosmological problem suggested in the present paper, 
I think that these results 
could also be relevant in other frameworks where scalar fields play a role. For the non 
perturbative solutions to Eq.(\ref{dlin}) of the kind $e^{\pm\phi}$, the gaussian potential is an 
UV stable fixed point. The same is true for those solutions of the kind (\ref{so}) when 
$\alpha>0$. This result is opposite to the perturbative one and its implications in 
particle physics models are certainly worth to explore. The scalar sector of the standard model is 
itself an open problem. It is not a priori clear whether the existence of these solutions
could have some relevance for the theory. I hope to come back to this issue in a future paper.
Work is in progress in this direction.


\begin{thebibliography}{12}
\bibitem{com1} As it is well known, being the $\phi^4$ operator marginal, we need to go up 
to the quadratic term to establish the flow of the quartic coupling constant $\lambda_4$.
\bibitem{com3} This is Eq.(18) of reference\cite{bo} once the trivial changes in the notations 
and the restriction to the $d=4$ case are made.
\end{thebibliography}
\end{document}